\documentclass[12pt]{article}
\usepackage[top=1in, bottom=1in, left=1in, right=1in]{geometry}
\usepackage[english]{babel}
\usepackage{atbegshi,cite}
\usepackage{amsmath,amssymb,amsbsy,amstext, amsthm, simplewick}
\usepackage{graphicx}
\usepackage{amsfonts}
\usepackage[small]{caption}
\usepackage{upgreek}
\usepackage[titletoc]{appendix}
\usepackage{setspace}
\usepackage{comment}
\usepackage{tensor}
\usepackage[usenames,dvipsnames,table]{xcolor}
\usepackage[colorlinks=true,urlcolor=blue
,anchorcolor=blue,citecolor=blue,filecolor=blue,linkcolor=blue,menucolor=blue
]{hyperref}

\def\pa{\partial}

\newcommand{\newc}{\newcommand}
\newc{\fpi}{f_{\pi}}
\newc{\etap}{\eta^{\prime}}
\newc{\llll}{\langle\lambda\lambda\rangle}
\newc{\FFd}{F^a\tilde F^a}
\newc{\qbar}{{\overline q}}
\newc{\TR}{{\rm Tr}}
\newc{\Kahler}{K\"ahler }
\newc{\Zbb}{{\mathbb Z}}
\newc{\Rt}{{{\mathbb R}^3}}
\newc{\Rf}{{{\mathbb R}^4}}
\newc{\Sth}{{{\mathbb S}^3}}
\newc{\SthSo}{{{\mathbb S}^3\times{\mathbb S}^1}}
\newc{\Stw}{{{\mathbb S}^2}}
\newc{\StwSo}{{{\mathbb S}^2\times{\mathbb S}^1}}
\newc{\So}{{{\mathbb S}^1}}
\newc{\zt}{{{\mathbb Z}_2}}
\newc{\RtSo}{{{\mathbb R}^3\times{\mathbb S}^1}}
\newc{\RfSo}{{{\mathbb R}^4\times{\mathbb S}^1}}
\newc{\scriminus}{{\cal I}^-}
\newc{\scriplus}{{\cal I}^+}
\newc{\mpl}{M_p}
\newc{\Ricci}{\mathcal{R}}
\newc{\bv}{\phi}
\newc{\calU}{{\cal U}}
\newc{\calK}{K}
\newc{\calUi}{{\cal U}^{-1}}
\newc{\calG}{{\cal G}}
\newc{\calH}{{\cal H}}
\newc{\calI}{{\cal I}}
\newc{\calL}{{\cal L}}
\newc{\calO}{{\cal O}}
\newc{\calQ}{{\cal Q}}
\newc{\calOb}{{\cal O}^\dagger}
\newc{\hphi}{{\hat\phi}}

\theoremstyle{plain}
\theoremstyle{plain} 
\theoremstyle{plain} 
\theoremstyle{plain}
\theoremstyle{plain}
\theoremstyle{plain}

\newc{\pd}[1]{{\color{Red}#1}}

\renewcommand{\title}[1]{{\Large\bf\flushleft{#1}}\vspace*{3ex}\\}
\renewcommand{\author}[2]{{\noindent\hspace*{2.5em}\large#1}
                     \footnote{Electronic mail: $\mathtt{#2}$}\\}

\newcommand{\beq}{\begin{equation}}
\newcommand{\eeq}{\end{equation}}

\begin{document}
\begin{titlepage}

\vskip 2.2cm

\begin{center}

{\Large \bf Vacuum Decay in Time-Dependent Backgrounds}
\vskip 1.2cm

{Patrick Draper\footnote{pdraper@illinois.edu}, Manthos Karydas\footnote{karydas2@illinois.edu}, and Hao Zhang\footnote{haoz17@illinois.edu} }
~\\
\vskip 1cm
{Illinois Center for Advanced Studies of the Universe \& Department of Physics,\\ University of Illinois, 1110 West Green St., Urbana IL 61801, U.S.A.}\\
\vspace{0.3cm}
\vskip 4pt

\vskip 1.5cm

\begin{abstract}
We develop semiclassical methods for studying bubble nucleation in models with parameters that vary slowly in time. Introducing a more general rotation of the time contour allows access to a larger set of final states, and typically a non-Euclidean rotation is necessary in order to find the most relevant tunneling solution. We work primarily with effective quantum mechanical models parametrizing tunneling along restricted trajectories in field theories, which are sufficient, for example, to study thin wall bubble nucleation. We also give one example of an exact instanton solution in a particular Kaluza-Klein cosmology where the circumference of the internal circle is changing in time.
\end{abstract}

\end{center}

\vskip 1.0 cm

\end{titlepage}

\setcounter{footnote}{0} 
\setcounter{page}{1}
\setcounter{section}{0} \setcounter{subsection}{0}
\setcounter{subsubsection}{0}
\setcounter{figure}{0}

\section{Introduction}

Almost fifty years ago, Coleman and collaborators developed powerful Euclidean path integral methods to compute vacuum decay rates in a semiclassical expansion~\cite{PhysRevD.15.2929,Callan:1977pt,Coleman:1980aw}. The phenomenon is of clear cosmological and existential relevance, and continues to drive interesting theoretical work, most importantly the development of new calculational tools~\cite{PhysRevD.95.085011,Espinosa:2018hue,Braden:2018tky,Guada:2020ihz,Ai:2020vhx,Lagnese:2021grb,Hayashi:2021kro,Matsui:2021oio,Croon:2021vtc,Ivanov:2022osf,deAlwis:2023gth}.

Euclidean path integral techniques are most straightforwardly applied to systems with time-independent couplings. However, if the tunneling field is coupled to other dynamically evolving fields, the tunneling potential may become effectively time dependent. In this paper we discuss generalizations of the semiclassical approach that can be used to compute tunneling probabilities  in systems with time dependent parameters.

In Sec.~\ref{secformulation} we describe a formulation of the semiclassical tunneling problem convenient for analyzing  scenarios with explicit time dependence. The time dependence must be adiabatic, so that it makes sense to focus on the instantaneous semiclassical false and true vacua. The main differences with the usual Euclidean approach are $(a)$ the introduction of a generalized time contour, and $(b)$ a reformulation of the variational problem, from fixed initial position and fixed energy, to fixed wavepacket-like initial and final states at fixed times. Actually, the wavepacket formulation is mainly useful for interpreting the results of computations where the initial energy is still provided: it maps the real and imaginary parts of the initial and final momenta in the semiclassical solution to simple features of the initial and final packet wavefunctions. 

In Sec.~\ref{secgamma} we examine the role of the rotation angle of the time contour in time-independent systems. We find that adjusting this contour allows us to identify different final states outside the potential barrier, all related to each other by classical time evolution. Another way to view this is that the time contour controls the time at which barrier penetration begins. In Sec.~\ref{sectime} we turn to time-dependent systems, beginning with a toy model that can be solved analytically. In this case the generalized time contour turns out to be essential: controlled solutions are only obtained if the time contour is not too close to the Euclidean axis.  Next we study thin-wall bubble nucleation in 3+1 dimensions with time dependent parameters. The leading correction to the semiclassical action can be obtained analytically, and solutions for the tunneling trajectories are easily obtained nnumerically. We describe how to make an optimal choice of continuation parameter depending on the path integral of interest.

In Sec.~\ref{sec:gravity} we discuss gravitational effects. We outline a general treatment, then consider some particular cases where approximations can make the problem more tractable. We find, for example, that the leading $\calO(H)$ contribution to the real part of the action vanishes for thin-wall nucleation on a fixed FRW background. We also describe, in Sec.~\ref{sec:KK}, an example of a Kaluza-Klein cosmology where a full bubble solution for the fields can be obtained. In this case the Lorentzian bubble metric is real, but the instanton is quasi-Euclidean, corresponding to a complex metric.

Sec.~\ref{sec:discussion} contains some comments on directions for further development.

\section{Semiclassical Formulation}
\label{secformulation}

In the usual semiclassical treatment of vacuum decay, one searches for saddle points of the Euclidean path integral that interpolate between the false vacuum and a bubble of true vacuum, or in quantum mechanics, between the false vacuum and a point outside the well. 

To get a different perspective on the problem, let us  consider  real-time quantum mechanical transition amplitudes of the form
\begin{align}
\langle \psi_f(R,&T_f) | \psi_i(R,T_i)\rangle = N\int dR_idR_f \int_{R(T_i)=R_i}^{R(T_f)=R_f} DR\, e^{iS/\hbar  }\,,\nonumber\\
&S=S_i(R_i)- S_f(R_f)+\int_{T_i}^{T_f} dt\, L(\partial_t R,R,t).
\label{basicamplitude}
\end{align}
 The initial and final states are given by wavefunctions parametrized as $\psi(R)=e^{iS_{i,f}(R)/\hbar}$, where $S_{i,f}$ are complex functions. In tunneling problems these wavefunctions should be localized inside and outside of a false vacuum well. We have written explicit factors of $\hbar $ to indicate that we are interested in states where both the real and imaginary parts of $S_{i,f}$ contain contributions of order one in the $\hbar$ expansion, like coherent states, but subsequently we will drop the $\hbar$s.   A similar starting point was used to study complexified path integral trajectories in~\cite{PhysRevA.83.012104}.
 
 Next, we perform a general rotation of the time contour, $t\rightarrow e^{i\gamma}\tau$. The  Euclidean continuation corresponds to $\gamma=-\pi/2$, but it turns out to be useful to keep $\gamma$ arbitrary. This technique was applied to the symmetric double well tunneling problem in~\cite{Cherman:2014sba}.\footnote{Among the interesting phenomena observed in~\cite{Cherman:2014sba} is that the ordinary Euclidean instanton, with a hyperbolic tangent profile, becomes a complex space-filling curve in the real-time limit $\gamma\rightarrow0$.  $\gamma$ will play a somewhat different role in the decay problems considered here.}
 
 After rotation of the time contour, the path integral appearing in the amplitude is 
\begin{align}
& \int dR_idR_f \int_{R(T_i)=R_i}^{R(T_f)=R_f} DR\, e^{-S_c}\,,\nonumber\\ 
 S_c=-iS_i(&R_i)+i S_f(R_f)-i \int_{T_i}^{T_f} d\tau\, L_c(\partial_\tau R,R,\tau)\, \nonumber\\
L_c&(\partial_\tau R,R,\tau)=e^{i\gamma} L(e^{-i\gamma}\partial_\tau R,R, e^{i\gamma}\tau).
\label{rotatedamplitude}
\end{align}
In the leading semiclassical approximation we are interested in solutions to the the bulk Euler-Lagrange equation, $\delta L_c/\delta R - \partial_\tau (\delta L_c/\delta \partial_\tau R)=0$. The equation of motion can be replaced by the conservation of energy condition if $L$ does not depend explicitly on $t$.

Finally, we relax the Dirichlet boundary conditions used in ordinary semiclassical treatments of the path integral. Since we have introduced general initial and final states, we can consider unrestricted boundary variations. Stationarity of the action then gives rise to the following boundary relations:
\begin{align}
p_i\equiv \frac{\delta L_c}{\delta \partial_{\tau} R}&\bigg|_{\tau =T_i} = S_i'(R_i)\nonumber\\
p_f\equiv \frac{\delta L_c}{\delta \partial_{\tau} R}&\bigg|_{\tau =T_f} = S_f'(R_f).
\label{initcond}
\end{align}
Eq.~(\ref{initcond}) will be important for understanding the various semiclassical tunneling solutions we obtain below. These  conditions relate the initial and final momenta of the semiclassical trajectories to the gradient of the wavefunction at the initial and final positions. A totally generic initial and final wavefunction, and a given solution $R(\tau)$, will not in general satisfy Eq.~(\ref{initcond}), and so will not provide a single saddle point approximation to the full path integral, including the initial and final states as described above. Our approach will be to find solutions to the bulk equations of motion first, and then use Eq.~(\ref{initcond}) to infer wavefunctions with consistent properties, such that those solutions provide genuine saddle points of the full path integral. These are generally weak constraints on the initial and final wavefunctions, only constraining their gradients at points, but we must determine them after the fact.

These self-consistency conditions have very natural physical interpretations. For example, the real part of the momentum at the semiclassical endpoints has to match the momentum of the initial and final state wavefunctions in the eikonal approximation.

At leading order in the semiclassical expansion, then, the task is to solve the equations of motion subject to the boundary conditions~(\ref{initcond}). 
The canonical variables in the solution are generally complex-valued. In addition to the $e^{i\gamma}$, $S_{i,f}$ are complex if, say, the  states defined on real $R_{i,f}$ are modeled as Gaussian wavepackets or some similar perturbative ground state. To estimate a total decay probability, one must also extremize over semiclassically inequivalent final states, and integrate over any zero modes.

Although this is an acceptable formulation of a tunneling problem, generally it is  not what is done in standard, time-independent tunneling problems. One instead solves a variational problem with fixed energy:
\begin{align}
E_E = E_{FV}
\label{typicalproblem}
\end{align}
where $E_E$ is the Euclidean energy and $E_{FV}=V(R_{FV})$ is the value of the potential in the classical false vacuum $R_{FV}$. Typically there is a solution to Eq.~(\ref{typicalproblem}) that starts at $R_{FV}$ and reaches the classical turning point $R=R_{\tiny TP}$ with zero momentum at some later time, which is determined by the solution. For example, in an inverted quartic double well model, 
\begin{align}
    L= \frac{1}{2} \dot R^2- R^2 + \lambda R^4,
\end{align}
the relevant zero-energy Euclidean solution is 
\begin{align}
    R = \frac{1}{\sqrt{\lambda}}{\rm sech}(\sqrt{2}\tau).
\end{align}
(In this case $R_{FV}$ and $R_{TP}$ are only reached asymptotically, but this is not important.)

From such a solution, we may work backward and identify  problems of the type~(\ref{basicamplitude})-(\ref{initcond}) to which it is also a solution. In a corresponding problem of the type (\ref{basicamplitude})-(\ref{initcond}), one needs a low-energy initial state, like the perturbative ground state wavefunction in the false vacuum, for which $S_i'(R_{\tiny FV})=0$, and similarly the final state can be some zero-momentum wavepacket peaked at $R_{\tiny TP}$. These requirements are typical of time-independent problems analytically continued to Euclidean time. 

We will see that other final states arise naturally in two more general situations: first, we consider time-independent problems with arbitrary time continuation parameters $\gamma$, and second, cases with explicit time-dependence in the action.
We solve the semiclassical equations of motion with initial energy as input, finding trajectories that end, typically, at some $R_f\neq R_{\tiny TP}$ with $p_f\neq 0$. From these solutions we may read off amplitudes of the form (\ref{basicamplitude}) for which they also provide saddle point trajectories. 
  In particular Eq.~(\ref{initcond}) provides relations between the complex semiclassical momenta and features of suitable wavepackets.

Let us pause and make a couple of tangential comments. First, in relating a complex-time solution to the original real-time amplitude~(\ref{basicamplitude}), there is a little more that needs to be said about the continuation of the parameters $T_{i,f}$ back to real time. We will discuss this issue below.  Second, we note that it is important {\emph{not}} to consider a path integral with fixed initial and final times and fixed initial and final values of $R$. Position eigenstates have completely uncertain momentum, and there will always be solutions, unrelated to tunneling, that have enough energy to summit the barrier classically. The fixed-energy formulation avoids these complications, as does the wavepacket formulation, as long as the gradients in the wavepacket are not too large.

\section{Time-Independent Tunneling with $t\rightarrow e^{i\gamma}\tau$}
\label{secgamma}

In this section we consider the general Wick rotation with parameter $\gamma$. For illustration, we work  with examples of the form
\begin{align}
L = -R_0 R^{p-1} \sqrt{1-(\partial_tR)^2 } + R^{p}.\label{model}
\end{align}
A large class of field theoretic vacuum decay processes are described by this effective Lagrangian at the leading semiclassical order, including thin-wall bubble nucleation in $d=p$ spatial dimensions, Schwinger pair production in a constant background electric field for $p=1$, and bubble of nothing decays for $p=d-1$. (Therefore  $R$ is assumed to be $\geq 0$, since  it  plays the role of a  radial collective coordinate.)  For example, consider a real scalar field in 3+1 dimensions with a double-well potential and a small symmetry breaking parameter $\epsilon$,
\begin{align}
    V(\phi)=\frac{1}{2}\lambda(\phi^2-a^2)^2 - (\epsilon/2a) \phi.
    \label{phi4theory}
\end{align}
It is well-known that the effective theory~(\ref{model}) with $p=3$ is obtained from a Born-Oppenheimer approximation to the Klein-Gordon field theory with potential~(\ref{phi4theory}). Let us briefly review the construction which will be useful to have in mind when we  include slow time dependence.

For small $\epsilon$ the semiclassical states relevant for tunneling in the theory~(\ref{phi4theory}) are large spherical domain walls. Neglecting the slow accelerations driven by curvature and the energy gap $\epsilon$, the equation of motion for the ``heavy" degrees of freedom is $\partial_t^2\phi -\partial_r^2\phi + 2\lambda(\phi^2-a^2)\phi=0$, for which the relevant solutions are
\begin{align}
    \phi\approx a \tanh\left[\frac{\sqrt{\lambda} a  (R+vt -r)}{\sqrt{1-v^2}}\right].
    \label{tanh}
\end{align} 
Here $R$ is a large radius, $v$ is the wall velocity,  the wall tension is given by $\sigma\propto \sqrt{\lambda} a^3$, and the factor of $\frac{1}{\sqrt{1-v^2}}$ accounts for length contraction of the wall thickness. To obtain an effective action for slow variations of $v$ driven by curvature and pressure, we replace $R+vt\rightarrow R(t)$, $v\rightarrow \dot R(t)$, and plug  these field configurations into the Klein-Gordon action. Upon so doing each term in the Lagrangian is found to be proportional to $f(r)\,{\rm sech}^4\left[\frac{\sqrt{\lambda} a  (R -r)}{\sqrt{1-\dot R^2}}\right]$. In the thin-wall limit $\frac{\sqrt{\lambda} a}{\sqrt{1-\dot R^2}} R \gg 1$ the ${\rm sech}^4$ is sharply peaked around $r=R$, so we may replace $f(r)\rightarrow f(R)$ and do the integral over space. One obtains an effective Lagrangian for the collective coordinate $R(t)$ which is, apart from a overall multiplicative constant, given by Eq.~(\ref{model}) with $p=3$ and $R_0=3\sigma/\epsilon.$ (We will be lax about such multiplicative constants, which do not affect the equations of motion and can be restored easily in the on-shell action.)

Since~(\ref{model}) does not depend explicitly on time, there is a conserved energy. After continuation $t\rightarrow e^{i\gamma}\tau$, the conserved energy is
\begin{align}
E=e^{i\gamma} \left(\frac{R_0 R^{p-1} }{\sqrt{1-e^{-2i\gamma}(\partial_\tau R)^2 }}-R^p\right).
\label{energy}
\end{align}
There is a false vacuum at $R_{FV}=0$ with $E_{FV}=0$, and the classical turning point is $R_{TP}=R_0$.  It is also informative to write the energy in terms of the momentum $P=e^{-i\gamma}R_{0}R^{p-1}\left(\pa_{\tau}R/\sqrt{1- e^{-2i\gamma}(\pa_{\tau}R)^2}\right)$,
\begin{align}
    E= e^{i\gamma}\left(\sqrt{P^2 + R_{0}^{2}R^{2(p-1)}}- R^{p}\right)\,.
    \label{hamiltoniangamma}
\end{align}

\begin{figure}[t!]
\centering
\includegraphics[width=0.55\textwidth]{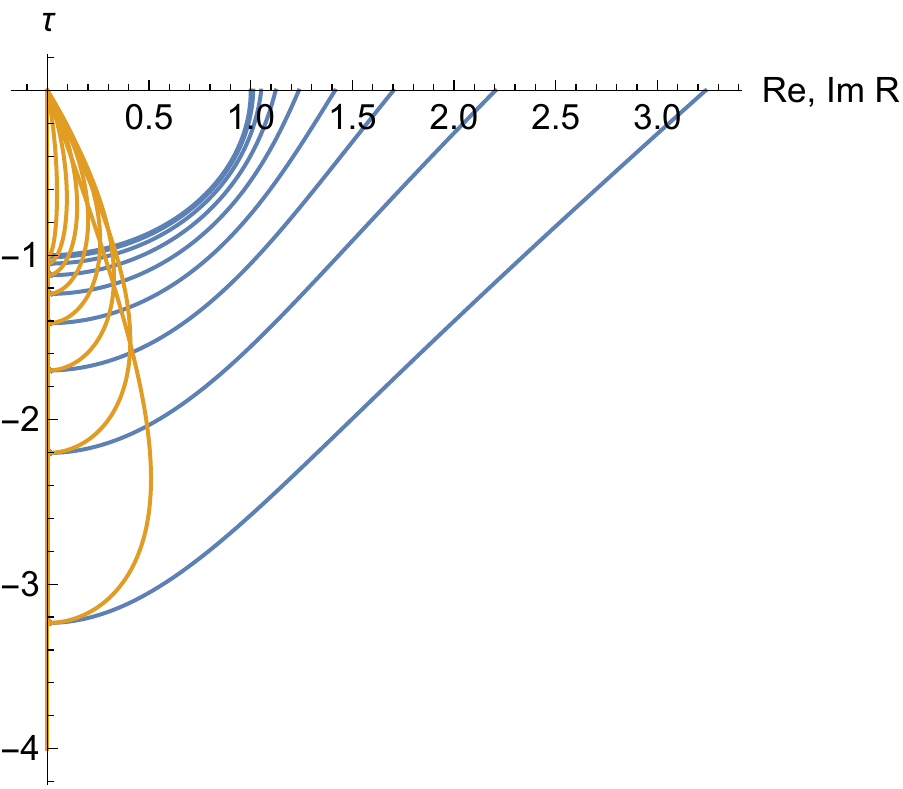}
\caption{Tunneling trajectories~(\ref{Rcirclegeneral}) with $R_0=1$ and a range of $\gamma$ between $-\frac{\pi}{2}$ (Euclidean) and $-\frac{\pi}{10}$. As $\gamma\rightarrow0^-$ the tunneling starts earlier and ends further outside the well, with finite classical momentum such that the total energy is conserved.
}
\label{fig:timeindeptraj}
\end{figure}

For $p>1$, $T_f=0$, and sufficiently early $T_i$, the solutions to Eq.~(\ref{typicalproblem}) relevant for tunneling are quite simple. In the Euclidean case $\gamma=-\pi/2$, they are 
\begin{align}
R = \Theta(\tau+R_0) \sqrt{R_0^2-\tau^2}.
\label{Rcircle}
\end{align}
The solution is real in the relevant range of $\tau$, and the nucleation point is $R(0)=R_0$, where the momentum vanishes. 
For other $\gamma$, $R$ is generally not real, so we seek a zero energy solution for which $R$ is real at $\tau=0$. Eq.~(\ref{Rcircle}) generalizes to
\begin{align}
R = \Theta(\tau-R_0\csc \gamma) \sqrt{R_0^2+(e^{i\gamma}\tau-R_0\cot\gamma)^2}.
\label{Rcirclegeneral}
\end{align}
The particle sits at the $R=0$ false vacuum until $\tau= \tau_0\equiv R_0\csc \gamma$, then moves in the complex $R$ plane to the nucleation point at $\tau=0$,
\begin{align}
R(0)=R_0|\csc \gamma|. 
\label{finalR}
\end{align} 
 Curiously, the nucleation point only coincides with the classical turning point in the Euclidean case. Examples are shown in Fig.~\ref{fig:timeindeptraj}.

We should say a word about the sense in which Eqs.~(\ref{Rcircle}),~(\ref{Rcirclegeneral}) are saddle points of the action. Since $\dot R$ is not defined at $\tau=\tau_0=R_0\csc\gamma$, we must give it a definition so that the trajectories are in the domain of the action functional. Any definition of $\dot R$ as a discontinuous function is sufficient and results in the same value of the action.  The discontinuity in $\dot R$  gives rise to a delta distribution in $\ddot R$, which could spoil the stationarity of the action under variations. However, for $p>1$ the delta function is multiplied by a positive power of $R$ in the action density, which vanishes at $\tau=R_0\csc \gamma$.

We have assumed that the initial time is early enough, $T_i < R_0\csc\gamma$, so that the instanton can ``fit."  Apart from this constraint, the action of the semiclassical solution is independent of $T_i$. The particle sits as long as it needs to in the false vacuum, and then at $\tau=R_0\csc \gamma$ it starts the jump to the nucleation point.

To better understand the solutions for $\gamma\neq -\pi/2$, we can ask what wavepacket problems of the form (\ref{basicamplitude})-(\ref{initcond}) they also solve. Let us examine the boundary conditions~(\ref{initcond}). The final momentum of the solutions~(\ref{Rcirclegeneral}) is
\begin{align}
p_f=-R_0^{p}|\csc\gamma|^{p-1}\cot\gamma.
\label{finalP}
\end{align}
In particular, it is real, and so  is related to the gradient of the phase of the final wavefunction by Eq.~(\ref{initcond}). For wavepackets the gradient of the phase is just the classical momentum $P$. 

Furthermore is easily seen that the {\emph{real-time}} energy, Eq.~(\ref{hamiltoniangamma}) with $\gamma=0$,
 vanishes for all final states labeled by the real positions $R\rightarrow R(0)$ given in Eq.~(\ref{finalR}) and  real momenta $P=p_{f}$ given in Eq.~(\ref{finalP}). This is because Eq.~(\ref{hamiltoniangamma}) only depends explicitly on $\gamma$ in an overall multiplicative factor, and Eq.~(\ref{initcond}) instructs us to map the semiclassical trajectory momentum $p_f$ and the semiclassical final state momentum $P$ to each other. The final states at $\tau=0$ correspond to different points in phase space, related by real-time classical  evolution.

We conclude that different values of $\gamma$ correspond to equivalent tunneling processes in an interesting way. 
 For $ -\pi < \gamma < -\pi/2$, the classical state is a particle somewhere to the right of the potential barrier, moving up the hill toward  the classical turning point. For $-\pi/2 < \gamma < 0$, the particle is moving away from the turning point. Classical evolution, either forward or backward in time, takes the particle to the state at rest at the turning point, which is the nucleation point for $\gamma=-\pi/2$. We may interpret this as meaning that different $\gamma$ (in the range $-\pi/2 < \gamma < 0$) correspond to different times at which a bubble nucleates at rest. The semiclassical path integral computes the amplitude to evolve from ``no bubble" at $T_i$ to ``some bubble" at $T_f$, and different $\gamma$ appear to naturally  account for the fact that we would also find a nonzero semiclassical amplitude, to nucleate a bubble with a smaller $p_f$, if we chose earlier $T_i$.

Consistent with this interpretation, all of the trajectories have the same on-shell value of the real part of the action. For example, for $p=3$, one finds
\begin{align}
    {\rm Re}\left[-i \int_{T_i}^{T_f} d\tau\, L_c(\partial_\tau R,R,\tau)\right] = \frac{\pi R_0^4}{16}
\end{align}
independent of $\gamma$.

According to Eq.~(\ref{rotatedamplitude}), the tunneling exponent  ${\rm Re}(S_c)$ also receives contributions from the initial and final state via ${\rm Im}(S_{i,f}(R_{i,f}))$. These factors can also be taken to be independent of $\gamma$. For example, the final states can be chosen as wavepackets related by translation in $R$.  (As long as the packets are localized away from the origin, the fact that $R$ is bounded from below is of no practical consequence.)

Finally, we must consider the analytic continuation of the amplitude back to real time. This is different from taking the limit $\gamma\rightarrow 0$ in the semiclassical solutions, because we have identified  solutions with different values of $\gamma$ as saddle points describing tunneling between different final states. Instead we want to continue the amplitude in the complex $T_i$-plane back to the real time axis, for {\emph{fixed}} final states.  Here we will be content with a plausibility argument. We have already seen that for fixed $\gamma$ the tunneling exponent is  independent of $T_i$ (if $|T_i|>R_0|\csc\gamma|$, so that the trajectory described by the instanton has enough time to complete.) Furthermore on physical grounds the real time decay rate is expected to be constant over timescales long compared to the perturbative timescales in the false vacuum and short compared to the lifetime. Therefore we make the plausible assumption that $T_i$ can be continued back to the real axis, trivially, for any  final state corresponding to some  $\gamma$. Furthermore we have seen that we obtain the same real part of the exponent, which controls the lifetime, for any $\gamma$.

\section{Time-Dependent Tunneling}
\label{sectime}

The formulations of the tunneling problem described above admit a straightforward generalization to time-dependent systems. 
We consider two explicit examples, a toy model and thin-wall bubble nucleation in 3+1 dimensions.

\subsection{Toy model}
The first example is simple enough that we can solve it more or less analytically:
\begin{align}
L = (\partial_tR)^2  -\left((R_0+v t) R - R^2\right).
\label{nrquadraticmodel}
\end{align}
As before the degree of freedom is constrained to $R\geq 0$, so that on the domain of $R$, the classical potential has a false vacuum at $R=0$. We assume $v>0$ so that the barrier is growing with time. We are free to choose the time parametrization so that the nucleation time is $T_f=0$, and we assume that $R_0$ is large enough so that $R_0+v T_i>0$. The model~(\ref{nrquadraticmodel}) has been scaled so that time is dimensionless and $v$ has the dimensions of $R$.

\begin{figure}[t!]
\centering
\includegraphics[width=0.55\textwidth]{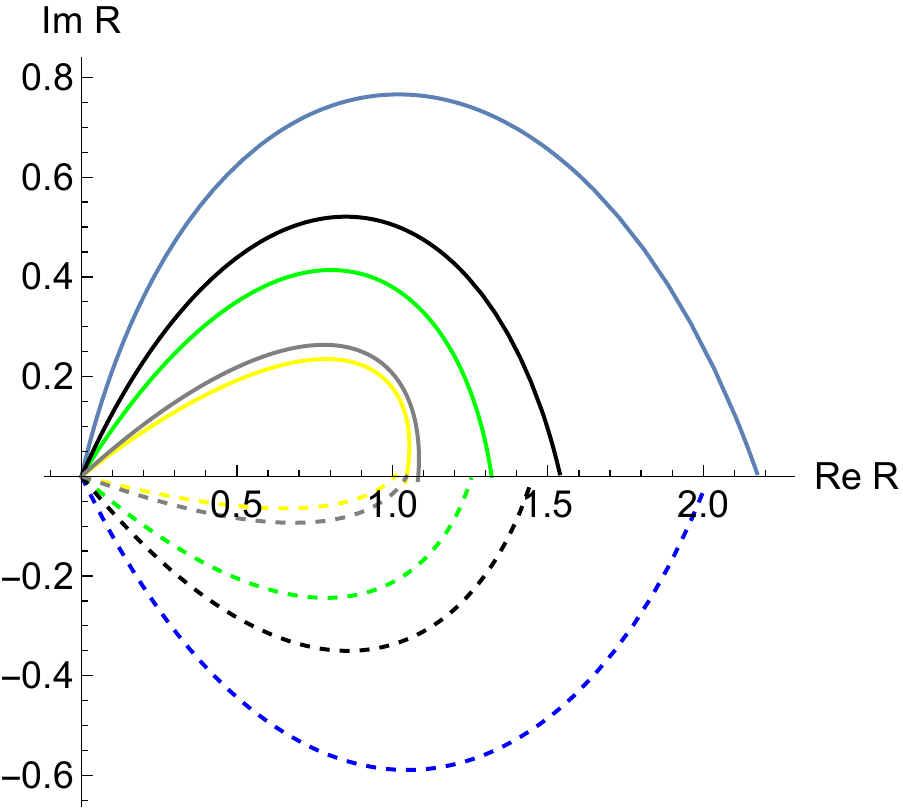}
\caption{Tunneling trajectories~(\ref{nrproblemsolution}) with $R_0=1$, $v=1/20$, and a range of $\gamma$ between $-\frac{\pi}{6}$ and $-\frac{2\pi}{3}$.
}
\label{toyorbits}
\end{figure}

We perform the same general continuation as previously, $t\rightarrow e^{i\gamma}\tau$, and seek solutions to the equations of motion  on which the classical Hamiltonian vanishes at some $\tau=\tau_0$.  
The tunneling solution is
\begin{align}
R(\tau)=\frac{1}{2}\left(R_0+e^{i \gamma } \tau  v\right)-\frac{1}{2}  \left(R_0+e^{i \gamma } \tau _0 v\right) \cosh \left(e^{i \gamma } \left(\tau -\tau _0\right)\right)
 -\frac{1}{2} v \sinh \left(e^{i
   \gamma } \left(\tau -\tau _0\right)\right).
   \label{nrproblemsolution}
\end{align}
Some examples are shown in the complex-$R$ plane in Fig.~\ref{toyorbits}.

To understand the decay process described by the solutions (\ref{nrproblemsolution}), we must determine the  tunneling time, the relevant initial and final states,  and the real part of the on-shell action. We consider each in turn.

Demanding that $R$ is real again at  $\tau=T_f=0$ determines the tunneling time $\tau_0$ via the somewhat messy relation
\begin{align}
R_0 \sin \left(\tau _0 \sin\gamma\right)& \sinh \left(\tau
   _0 \cos\gamma\right)=v\bigg(\sin \left(\tau _0 \sin\gamma\right) \cosh \left(\tau _0 \cos (\gamma
   )\right)\nonumber\\
   &-\tau _0 \left[\sin\gamma \left(\cos \left(\tau _0 \sin \gamma
   \right)\right) \cosh \left(\tau _0 \cos\gamma\right)+\cos\gamma \sin
   \left(\tau _0 \sin\gamma\right) \sinh \left(\tau _0 \cos \gamma
   \right)\right]\bigg)
 \label{imR0}
\end{align}
For nonzero $v$ and  Euclidean time, $\gamma=-\pi/2$, (\ref{imR0}) reduces to $\tan(\tau_0)=\tau_0$, which has no nontrivial solution.  Conversely it is  satisfied by any $\tau_0$ if $v=0$ and $\gamma=-\pi/2$. We solve it perturbatively in small $v$:
\begin{align}
    \tau_0 = \pi  \csc \gamma \bigg(1-\frac{v}{R_0}\text{coth}(\pi  \cot
   \gamma ) + \calO(v^2)\bigg)
   \label{pertTfsol}
\end{align}
where the higher order coefficients are lengthy but easily calculable. Near $\gamma=-\pi/2$, the solution (\ref{pertTfsol}) behaves as
\begin{align}
    \tau_0\sim-\left(\pi+(1/6 + \pi^2)\frac{v^2}{\pi R_0^2}\right)  -\frac{v}{
    R_0}\left(\gamma +\frac{\pi }{2}\right)^{-1} +  \frac{v^2}{ \pi   R_0^2}\left(\gamma +\frac{\pi }{2}\right)^{-2}+\dots
\end{align}
so the small-$v$ expansion is only controlled for 
\begin{align}
    |\gamma+\pi/2|\gg \frac{v}{\pi R_0}.
    \label{gammaexpansion}
\end{align}
It is also possible to obtain $\tau_0$ as an expansion in $\gamma+\pi/2$ for fixed nonzero $v$, but one finds that $R_f$ is small, of order $\gamma+\pi/2$, and the solutions typically do not describe tunneling out of the well. In other words, we find the surprising property that in this problem we must not work too close to the imaginary time axis in order to obtain classical solutions relevant for tunneling.

Now we examine the initial and final states.  To describe tunneling between two fixed times, we set $R=0$ for $T_i\leq \tau\leq\tau_0$, and $R$ is given by Eq.~(\ref{nrproblemsolution}) for $\tau>\tau_0$. 
Despite its piecewise nature, it is easy to check that this trajectory is still a genuine stationary point of the continued action.\footnote{Actually, by constructing a model simple enough to solve analytically, we have inadvertently complicated this step, because the potential is not differentiable at the origin. Its derivative must be defined to be discontinuous, set to zero for all $t$ at $R=0$, and $R_0+v t-2R$ for $R>0$. This is necessary for the vacuum solution  $R=0$ to exist. This technical complication will not arise in the bubble nucleation problem considered in the next section.} As for the final state, we find an $\calO(v)$ correction to the nucleation point,
\begin{align}
R_f(0) = R_0 + \frac{1}{2} R_0 (\cosh (\pi  \cot \gamma )-1)-\frac{1}{2} v\sinh (\pi  \cot \gamma ) + \dots.
\end{align}
The $v=0$ contribution exhibits the same behavior as the time-independent model of the previous section: as $\gamma\rightarrow 0^-$ it diverges, indicating nucleation far outside the barrier. On the other hand, near the Euclidean time axis (keeping leading terms of order $v^2$),
\begin{align}
R_f(0) = R_0 + \frac{(5\pi^{2}-2)v^{2}}{24 R_{0}} + \frac{1}{2} v\pi (\gamma+\pi/2) - \frac{v^2}{4 R_0} (\gamma+\pi/2)^{-2}+\dots.
\end{align}
The third term is negative if $\gamma<-\pi/2$ and the fourth term is negative definite. The second term is positive but for $|\gamma+ \pi/2|\gg v/(\pi R_{0})$ the third term is larger in absolute value so the sum of two is also negative. Within the small-$v$ approximation, all of these terms must be much smaller in magnitude than $R_0$. So a novel feature of time dependent systems, which do not conserve energy, is that it is possible to obtain a nucleation point  less than the classical turning point at the nucleation time. However, in the present model this effect is atypical and marginal in the small-$v$ expansion.

The final momenta of the trajectories are
\begin{align}
    {\rm Re}(p_f) &= -R_0\sinh(\pi\cot \gamma ) +\calO(v) \nonumber\\ 
    {\rm Im}(p_f) &= \pi v\, {\rm csch}(\pi\cot \gamma )+\calO(v^2).
\end{align}
As before, the final state is characterized by a nonzero classical momentum ${\rm Re}(p_f)$ for general $\gamma$. The new feature is a nonzero imaginary component of $p_f$ for nonzero $v$. For typical $\gamma$ not too close to the Euclidean limit, $ {\rm Im}(p_f) $ is exponentially small, of order  $ v e^{-\pi/\gamma}$.  ${\rm Im}(p_f)$ is related to the gradient of the final-state wavepacket at the nucleation point, in problems of the type~(\ref{basicamplitude}), by Eq.~(\ref{initcond}). The imaginary part of $S_f$ and its gradient, for a wavepacket peaked at some $R_\star$, satisfy
\begin{align} 
{\rm Im}(S_f)  &= (R_f-R_\star)^2/2\sigma^2,\nonumber\\
{\rm Im}(S_f')  &= (R_f-R_\star)/\sigma^2.
\end{align}
A typical size of $\sigma$ is $\lesssim {\rm Re}(p_f^{-1})$. Consequently the typical distance from $R_f$ to the peak of the wavepacket, for $|\gamma|\lesssim \calO(1)$, is also exponentially small in $\pi/\gamma$. So to a good approximation we can still think of the final state as being peaked at $R_f$, if $\gamma$ is not too close to the Euclidean limit, and the state possesses finite classical momentum. 

In the opposite limit, close to Euclidean time (while maintaining the bound~(\ref{gammaexpansion})), the classical momentum of the final state  ${\rm Re}(p_f)$ goes to zero like $\gamma+\pi/2$: the particle nucleates approximately at rest. The gradient of the final state wavefunction at the nucleation point grows like $(\gamma+\pi/2)^{-1}$: the stationary point of the wavepacket problem balances a decrease in the tunneling distance against a decrease in the amplitude of the final state wavefunction.

The decay amplitude is governed by the real part of the on-shell action,
\begin{align}
    {\rm Re }\, S_c = {\rm Im  }\, S_i(0) - {\rm Im  }\, S_f(R_f) + \frac{\pi  R_0^2}{4 }+\frac{1}{2} \pi^2  R_0 v  \cot (\gamma )+\calO(v^2).
    \label{ReScnr}
\end{align}
The first two terms depend on the wavepackets at the starting and nucleation points. Since typical semiclassical solutions permit $R=0$ and $R=R_f$ to lie close to the packet peaks,  these terms may be neglected. (There is a contribution from the wavefunction normalizations, but it is of higher order in the $\hbar$ expansion.) The rest of~(\ref{ReScnr}) corresponds to the real part of the action ${\rm Re}\,(-i \int L_c d\tau)$ evaluated on the semiclassical trajectory. 

The leading term in the $v$ expansion, $\pi R_0^2/4$, corresponds to the time-independent semiclassical action. The $\calO(v)$ correction can actually be computed just from the $\calO(v)$ term in the action, evaluated on the $v=0$ solution. This is not completely obvious because the perturbation also changes the solution at the initial and final times, as well as the initial time. The full perturbation to the on-shell action at order $v$ takes the form:
\begin{align}
    \Delta S_c = v\left[ \left(i e^{2i\gamma}\int_{\tau_{0}}^0 d\tau \  \tau R\right)+\left(iL_c(R,\tau=\tau_0)\partial_v\tau_0\right)  + \left(i p\partial_v R\right)\bigg|_{\tau=\tau_0}^{\tau=0}\right]\bigg|_{v=0}
    \label{actioncorr}
\end{align}
The second term arises from the $\calO(v)$ correction to the initial time $\tau_0$. It vanishes because $L_c=0$ at the initial time for $v=0$. The third term involves the momentum $p=2e^{-i\gamma}\partial_\tau R$. It arises from the integration by parts in the kinetic term when the on-shell action is varied with respect to $v$. Physically, it translates the initial and final $R$ by $\delta R=v\partial_v R$.  The $\tau=\tau_0$ contribution from this term vanishes because $\partial_\tau R$  vanishes at the initial time. The contribution from $\tau=0$ is nonzero, but purely imaginary because $p_f$ is real for $v=0$. (Of course, these boundary variations must also cancel if Eq.~(\ref{initcond}) is satisfied.) Thus the first, ``bulk" term in~(\ref{actioncorr}) accounts for the full $\calO(v)$ term in (\ref{ReScnr}), as can be  verified explicitly,
\begin{align}
    \Delta {\rm Re} S_c|_{\calO(v)}=\frac{1}{2} \pi^2  R_0 v  \cot (\gamma ).
\end{align}

Physically, the $\calO(v)$ correction to the tunneling exponent is sourced by several effects. First, the tunneling takes place at earlier times, when the potential barrier was smaller by an amount of order $R_0 v \csc\gamma$. (Recall that we assume $v>0$, so that the barrier is growing in time; it is straightforward to reinterpret for negative $v$.) This effect decreases the action. Second, the tunneling duration $|\tau_0|$ (cf. Eq.~(\ref{pertTfsol})) has a  $\calO(v)$ correction, which also affects the action at $\calO(v)$. The duration is longer for $\gamma < -\pi/2$, increasing the action, and shorter for $\gamma > -\pi/2$, further decreasing it. The net $\calO(v)$ shift in the action is positive for $\gamma < -\pi/2$ and negative for $\gamma > -\pi/2$. 

The magnitude of the $\calO(v)$ contribution appears to decrease (increase) without bound as $\gamma\rightarrow 0^-$ ($\gamma\rightarrow -\pi^+$), in both cases because $\tau_0$ is diverging to $-\infty$. This reflects two related facts. First the $v$ expansion of the action breaks down at large $\tau_0\sim R_0/v$, and so (\ref{ReScnr}) is not reliable unless $\gamma \ll -\pi v/R_0$ and $\gamma \gg \pi+\pi v/R_0$. Then the correction is under control and the real part of the action is positive definite. Second, we should incorporate the bounds $R_0/v>|T_i|>|\tau_0|$, so that the tunneling processes have time to complete, and the barrier is already present at the initial time $T_i$ when the system was known to be in its vacuum state. This implies the constraint $\pi v|\csc \gamma|<R_0 $ for small $v$, which limits $\gamma$ in the same way.

Apart from these bounds, what $\gamma$ should we pick? The smallest action arises when tunneling starts right away, so that $T_i=\tau_0$ and the barrier is minimized. For small $v$ this means we have $\pi \csc\gamma \approx T_i$. Then we find
\begin{align}
    \Delta {\rm Re} S_c|_{\calO(v),{\rm max}}=-\frac{1}{2} \pi  R_0 v  \sqrt{T_i^2-\pi^2}
    \label{DeltaStoymodel}
\end{align}
for any $-\pi>T_i > -R_0/v.$ The upper bound on $T_i$ requires that at least one instanton ``fits" before the nucleation time $T_f=0$, and the lower bound on $T_i$ corresponds to the time when the barrier vanishes; note that here the correction to the action is of the same order as the leading term and of opposite sign.

The correction~(\ref{DeltaStoymodel}) is straightforward to understand. Part of the effect is translating the coupling $R_0$ appearing in the leading time independent action, $\frac{\pi }{4}R_0^2$, back to the starting time, $R_0\rightarrow R_0-v T_i$, at first order in $v$. This effect is dominant if $T_i\gg\pi$. The rest of the correction is the leading effect of time dependence while the particle is moving under the barrier. It vanishes in the Euclidean case, $T_i=\pi$, because here there is an extra $-i\tau$ in the $\calO(v)$ Lagrangian, which makes the contribution purely imaginary.

As we found in the time-independent case, generic values of $\gamma$ correspond to bubbles in motion at $T_f=0$. As in the time-independent case it is natural to conclude that the solution is telling us that nucleation has effectively already occurred at an early time: we can evolve the characteristic position and momentum of a suitable wavepacket backward in real time to a state of smaller momentum closer to the classical turning point at that time. Our path integral only ``asks" about events at $T_f$, but it ``knows" about decays that occured earlier.

Holding $\gamma$ fixed, the on-shell action Eq.~(\ref{ReScnr}) is independent of $T_i$ as long as the instanton fits. Therefore we may trivially continue the semiclassical amplitude $e^{-S_c}$ back to real time, $T_{i,f} \rightarrow e^{-i\gamma} T_{i,f}$. This continuation only acts nontrivially on $T_i$, since $T_f=0$.

We conclude that the solutions (\ref{nrproblemsolution}) provide a semiclassical approximation to the probability amplitude~(\ref{basicamplitude}), with some low-energy initial state peaked at $R=0$ at a specified initial time, and a  final wavepacket localized near $R_f$ at $T_f=0$, with properties related to the real and imaginary parts of the final momentum by Eq.~(\ref{initcond}). As before, we have avoided the more difficult problem of solving the boundary value problem (\ref{initcond}) with specified $S_{i,f}$ directly, and instead solve the initial value problem with specified $R$ and $\dot R$. The price is that we did not know exactly what family of wavepacket transition amplitudes we  solved until we computed the total tunneling time, the nucleation point, and final momentum of the solution.

\subsection{Thin-wall bubble nucleation}

Our second example is the 3+1 dimensional  thin-wall bubble nucleation model. We consider the case of a time-dependent domain wall tension proportional to $\sigma(t)$ and vacuum energy splitting $\epsilon(t)$ (as before, geometric factors have been absorbed to reduce clutter):
\begin{align}
L = -\sigma(t) R^{2} \sqrt{1-(\partial_tR)^2 } + \epsilon(t) R^{3}.
\label{timedepthinwallmodel}
\end{align}
We  assume that $\sigma(t),\,\epsilon(t)$ are positive for all relevant times.

Before analyzing the tunneling processes in this model, let us discuss how ~(\ref{timedepthinwallmodel}) might arise as an effective action in a field theory  coupled to a dynamical  sector. For example, we can weakly couple the model (\ref{phi4theory}) to another scalar $\chi$ rolling in a shallow potential,
\begin{align}
V(\phi,\xi)=\frac{1}{2}\lambda(\phi^2-a^2)^2 - \frac{\epsilon_0}{2a}(1+g \chi/a) \phi + V(\chi) .
    \label{phichitheory}
\end{align}
Suppose that  on the timescales of interest we can linearize $V(\chi)\approx c/a \chi$. The specifications of ``weak coupling" and ``shallow potential" mean we take the hierarchy $\lambda a^4 \gg (c \chi/a, \epsilon_0)  \gg g \epsilon_0 \chi/a $ over all relevant $\chi$. Then we can add the effects of $\chi$ to the Born-Oppenheimer treatment described in Sec.~\ref{secgamma}. In the first step, we solve the approximate ``heavy" dynamics, obtaining the  profile (\ref{tanh}) at zeroth order in  $\epsilon_0$, $c$, and $1/R^2$. Subsequently we take the thin-wall limit.  Then we obtain the ``light" dynamics, finding a homogeneous rolling solution for $\chi$ at leading order in the hierarchy $c\gg g\epsilon_0$,
\begin{align}
    \chi=\chi_0+ vt+\frac{1}{2} (c/a) t^2,
    \label{linearchi}
\end{align} 
for some $\chi_0$ and $v$. 
Evaluating the Lagrangian on the thin-wall profile and  Eq.~(\ref{linearchi}) and integrating over space, we obtain the quantum mechanical model~(\ref{timedepthinwallmodel}) with constant $\sigma = 16\pi \sqrt{\lambda}a^3 /3$ and $\epsilon(t)=4\pi/3 [\epsilon_0+(g/a)(\chi_0+vt+\frac{1}{2} c t^2)].$ 

With other weak couplings and potentials we may obtain arbitrary functions $\sigma(t)$ and $\epsilon(t)$, and possibly more general couplings. It is also straightforward to include backreaction of the $\phi$ profile on $\chi$ in the matching of the effective parameters, although this is higher order in $g$ and may be subleading to corrections to the thin-wall limit.

Now let us study the tunneling solutions of (\ref{timedepthinwallmodel}), along the same lines as our analysis of the toy model. 
Unfortunately, in the present model, typically  an analytic solution is no longer available. However, it is straightforward to construct numerical solutions, and the correction to the tunneling exponent may be obtained analytically when the parameters vary slowly.

To construct numerical solutions we require initial conditions.
At early times the relevant nontrivial solutions behave as
\begin{align}
R &= c\sqrt{\tau-\tau_0}\left(1+\calO(\tau-\tau_0)\right)\,,\nonumber\\
c &= ie^{i\gamma/2}\sqrt{\frac{6(\sigma_0+e^{i\gamma}\tau_0 v_\sigma)}{3i(\epsilon_0+e^{i\gamma}\tau_0v_\epsilon)+v_\sigma}}
\label{initdata}
\end{align}
where $v_{\sigma,\epsilon}$ are defined by the early time expansions
\begin{align}
\sigma (e^{i\gamma} \tau) &= \sigma_0 + e^{i\gamma} v_\sigma (\tau-\tau_0) + \dots,\nonumber\\
\epsilon(e^{i\gamma} \tau) &= \epsilon_0 + e^{i\gamma} v_\epsilon (\tau-\tau_0) + \dots.
\end{align}
Using (\ref{initdata}) as initial data, we may integrate the equations of motion   with specified $\sigma(t)$, $\epsilon(t)$ until $R$ becomes real again, which defines a nucleation point $R_f$. We may adjust $\tau_0$ so that the nucleation time is  $T_f=0$. Call such a solution $\hat R(\tau)$. As before, we can also attach an arbitrary period of the solution $R=0$ to the beginning of the solution $\hat R(\tau)$ to obtain a tunneling solution of longer total duration,
\begin{align}
R(\tau) = \theta(\tau)\hat R(\tau)\,;\;\;\;\; T_i<\tau<0.
\label{generalsolution}
\end{align}

\begin{figure}[t!]
\centering
\includegraphics[width=0.48\textwidth]{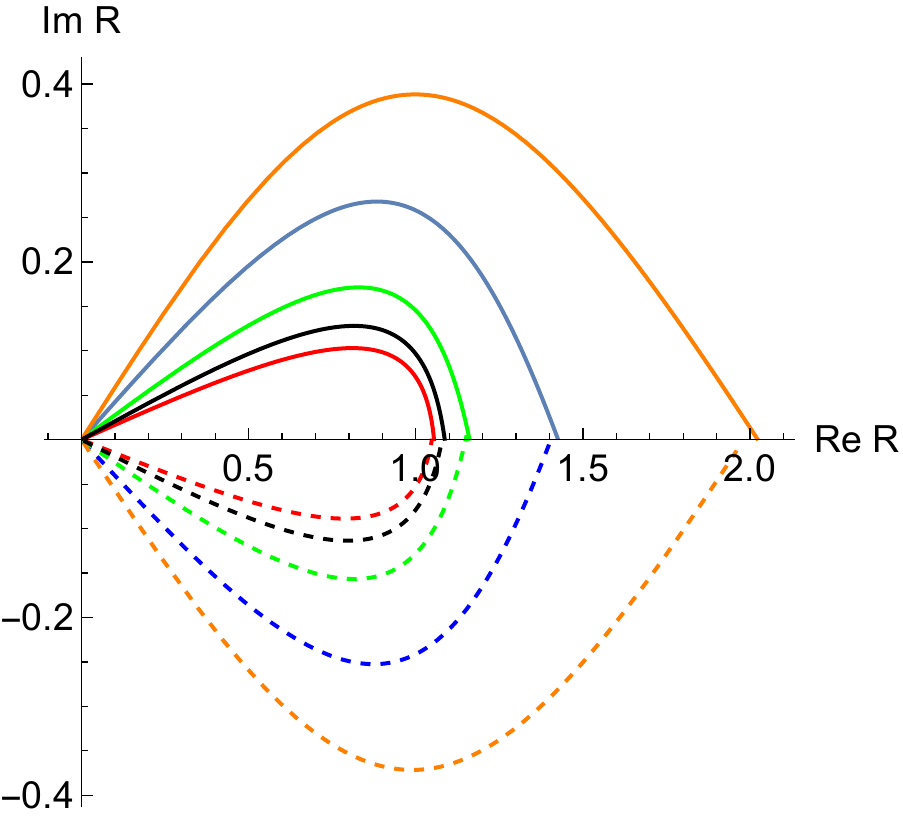}~~
\includegraphics[width=0.48\textwidth]{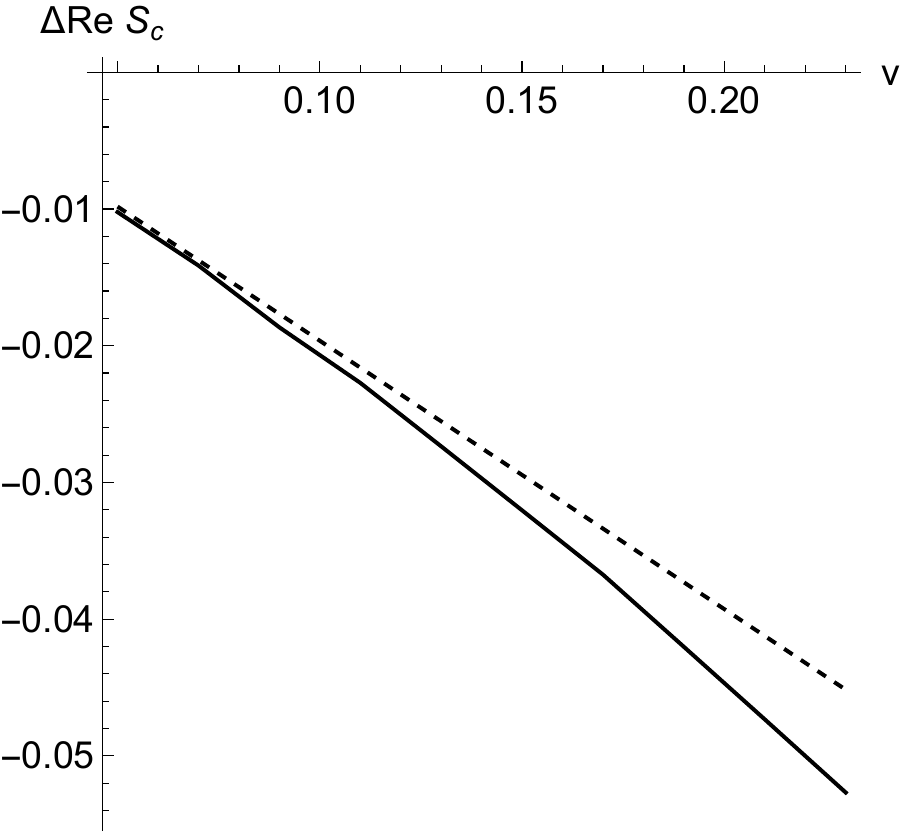}
\caption{Left: Tunneling trajectories with  $R_0=1$, $R_0v_\sigma/\sigma =R_0v_\epsilon/\epsilon =\frac{1}{20}$, and a range of $\gamma$ between $-\pi/6$ and $-2\pi/3$. 
Right: the $v$-dependent correction to the real part of the on-shell action (solid) compared to the analytic  result linearized in $v$ (dashed, Eq.~(\ref{linearvSc})), for $\gamma=-\pi/4$ and a range of $v$. (Here $\Delta {\rm Re}\ S_c$ is normalized to $R_0^3 \sigma_0$ which controls the leading $v$-independent term in the  on-shell action.)
}
\label{fig:bubblemodel}
\end{figure}

In Fig.~\ref{fig:bubblemodel} we show example trajectories in a model with linear time dependence in $\sigma$ and $\epsilon$, such that the barrier height grows linearly in time. 

In some cases, we can compute corrections to the tunneling exponent without knowledge of the full solution. Suppose that the parameters vary slowly enough that the linearizations~(\ref{initdata}) hold to a good approximation between $\tau=\tau_0$ and $\tau=0$. The $\calO(v_\epsilon, v_\sigma)$ corrections to the on-shell action are given by the $\calO(v_\epsilon, v_\sigma)$ terms in the action evaluated on the $v_\epsilon=v_\sigma=0$ solution, Eq.~(\ref{Rcirclegeneral}).\footnote{The reason is essentially the same as in the analysis of the toy model, Eq.~(\ref{actioncorr}). In the bulk time window the corrections to the solution itself only affect the action at $\calO(v^2)$. The various boundary contributions either vanish because $R$ vanishes at the initial time, or are purely imgainary because the final momentum for $v_{\epsilon,\sigma}=0$ is real, cf. Eq.~(\ref{finalP}). Thus the entire $\calO(v)$ correction to the real part of the action comes from the bulk $\calO(v)$ term evaluated on the $v=0$ solution (\ref{Rcirclegeneral}).}   We obtain:
\begin{align}
    \Delta {\rm Re}~S_c|_{\calO(v)} = \frac{\pi}{4}\cot(\gamma)R_0^4\left(v_\sigma-\frac{3}{4}R_0 v_\epsilon\right).
\end{align}
In Fig.~\ref{fig:bubblemodel} we compare the linear approximation to the full numerical result in some examples, finding agreement consistent with $\calO(v^2)$ corrections.

As in the toy model studied in the previous section, we can map these solutions to wavepacket transition amplitudes as in~(\ref{basicamplitude}). The solution begins at  $R=0$ at an  initial   $\tau = T_i$. The particle remains at rest at the origin until $\tau=\tau_0$, then moves to the nucleation point $R_f$ at $T_f = 0$, where it possesses some final momentum $p_f$. The action is independent of $T_i$ as long as  $T_i < \tau_0$, and so we  trivially continue the semiclassical amplitude $e^{-S_c}$ back to real time, $T_{i} \rightarrow e^{-i\gamma} T_{i}$. 

For fixed $T_i$, and assuming the barrier is growing monotonically sufficiently rapidly, the largest tunneling amplitude corresponds to $\gamma$ such that $\tau_0=T_i$. For slow time dependence, this implies the relation $R_0\csc \gamma\approx T_i$, and the  correction to the tunneling exponent
\begin{align}
    \Delta {\rm Re}~S_c|_{\calO(v),\rm max} = -\frac{\pi}{4}\sqrt{T_i^2-R_0^2}R_0^3\left(v_\sigma-\frac{3}{4}R_0 v_\epsilon\right).
    \label{linearvSc}
\end{align}
Again the correction~(\ref{linearvSc}) is easy to interpret. Part of the effect is translating the couplings $\sigma_0$ and $\epsilon_0$ appearing in the leading time independent action, $\frac{\pi}{16}\sigma_0^4/\epsilon_0^3$, back to the starting time $T_i$ in the linearized approximation. This effect is dominant if $T_i\gg R_0$. The rest of the correction is the leading effect of time dependence while the particle is moving under the barrier. The former correction is captured by the replacements $\sigma_0\rightarrow \sigma(t)$, $\epsilon_0\rightarrow \epsilon(t)$, in the leading on-shell action at the linearized level. However, we see that this is an incomplete accounting of the $\calO(v)$ terms, and corrections that arise from time dependence during the main part of barrier penetration are obtained in the more systematic analysis. The correction vanishes for the Euclidean case $T_i=R_0$ because the shift in the action is purely imaginary; this is a consequence of linearizing the time dependence, and nonzero corrections should be expected even in the Euclidean case if the time dependence is nonlinear.

More generally, the leading correction to the tunneling exponent should be obtained from the leading time-dependent perturbation to the time-independent action, integrated over the time-independent trajectories, and minimizing the real part over $\gamma$.

\section{Coupling to gravity}
\label{sec:gravity}

In principle, the thin-wall bubble nucleation formalism used above may be extended to include gravitational effects. In static cases, like the thin-wall limit of~(\ref{phi4theory}), one can construct a suitable membrane effective action. We will discuss this case below and generalize slightly an effective action constructed by Visser~\cite{Visser:1990wi}. The same action was used in \cite{1996PhyEs...9..556A} to compute decay rates in special cases using euclidean methods. In the presence of time-dependent driving fields, however, the construction of a complete membrane effective action is more cumbersome, and it is convenient to adopt some approximations. In one class of problems the gravitational field is the primary source of time dependence, and in favorable circumstances we can neglect the backreaction of the bubble on the gravitational field. In another class of problems the backreaction of the bubble on the field is accounted for, while the effects of time-dependent sources on the gravitational field can be neglected. We consider each case in turn.

As an example of the first type, suppose that we place the model~(\ref{phi4theory}) on a rigid FRW background, $ds^2 = -dt^2 +a(t)^2 dx_i^2$. Slow time dependence corresponds to tunneling processes that proceed quickly compared to the Hubble time $a/\dot a$. Suppose that we are interested in tunneling between times $t_i$ and $t_f$, and we normalize the scale factor to 1 at $t_f$. Then we may set 
    $a \approx 1+H (t-t_f)$
where $H$ is the Hubble parameter at $t_f$. To match to our framework in which tunneling ends at time zero, we shift the time coordinate, $t \rightarrow  t+t_f$. Then
\begin{align}
    a=1+Ht
\end{align}
and the Klein-Gordon action expanded to $\calO(H)$ is 
\begin{align}
    S\approx \int d^4x \left[\calL_{flat} + Ht (3\calL_{flat} + (\partial_i\phi)^2)\right].
\end{align}
The corresponding effective Lagrangian is 
\begin{align}
L = -(1+3Ht) \sigma R^{2} \sqrt{1-(\partial_tR)^2 } + (1+3Ht)\epsilon R^{3} + \frac{\sigma Ht R^2}{\sqrt{1-(\partial_tR)^2 }}
\label{timedepthinwallmodelFRW}
\end{align}
with $\sigma$ and $\epsilon$ given by their flat-space values, $\sigma = 16\pi \sqrt{\lambda}a^3 /3$ and $\epsilon=4 \pi \epsilon_0/3 $. We see that in addition to time-dependent $\epsilon$ and $\sigma$, as in the model~(\ref{timedepthinwallmodel}), there is also a new term. 

The real part of the $\calO(H)$ correction to the on-shell action, obtained from the $\calO(H)$ term in Eq.~(\ref{timedepthinwallmodelFRW}), is found to vanish. In typical FRW spacetimes the acceleration is of order $H^2$, so if there are $\calO(H^2)$ corrections they must be found retaining both the acceleration and terms of order $H^2$ in the action and terms of order $H$ in the solution. This problem can be studied numerically, but we will not pursue it further here.

Instead we consider another class of problems, where the gravitational field is static, but some gravitational backreaction from the bubble is incorporated. We start with the fully static problem.  The Einstein-Hilbert action coupled to a brane is \footnote{Other boundary terms besides ${\cal T}$ will be neglected since they do not affect effective action.}
\beq
\label{EH_action_bubble}
S^{EHGH}_{1,2}= \frac{1}{16\pi G_{N}}\int_{{\cal M}_{1,2}}\sqrt{|g|}d^{4}x \left({\cal R}_{1,2}-2\Lambda_{1,2}\right) + \frac{1}{8\pi G_{N}}\int_{{\cal T}}\sqrt{|h|}d^{3}{y}K_{1,2}- \mu \int_{{\cal T}} \sqrt{|h|}d^{3}y\,,
\eeq
where $K$ is the trace of the extrinsic curvature tensor and ${\cal T}$ denotes the brane worldvolume. The effective action for the location of the brane $R$ is obtained by integrating out the bulk fields. We assume that the relevant true and false vacuum spacetimes are approximately de Sitter in the static slicing. The appropriate action is:
\beq
\label{Bubble_effective_action_lapse}
\begin{split}
S^{\text{eff}}_{\text{bubble}}= \frac{4\pi}{8\pi G_{N}} \int d\lambda& \bigg[2 R\dot R\, \text{sinh}^{-1}\left(\frac{\dot R}{\sigma_{\text{dS}_F}\sqrt{N^{2}f_{\text{dS}_F}}}\right) -2 R \dot R\,\text{sinh}^{-1}\left(\frac{\dot R}{\sigma_{\text{dS}_T}\sqrt{N^{2}f_{\text{dS}_T}}}\right)\\&
+ 2\sigma_{\text{dS}_T}R\sqrt{f_{\text{dS}_T}N^{2}+ \dot R^2} -2 \sigma_{\text{dS}_F}R\sqrt{f_{\text{dS}_F}N^2 + \dot R^2}- 8\pi G_{N}\mu N R^2\bigg]\,,
\end{split}
\eeq
where $\dot R= \frac{dr}{d\lambda}$ and the subscripts $T$ and $F$ refer to the true and false vacua with $f_{dS_{T,F}}=1- R^2/L_{T,F}^2$. In general the value of each $\sigma_{dS_{T,F}}$ could be $\pm 1$. This action slightly generalizes that of Visser~\cite{Visser:1990wi} (to which we refer for a more detailed derivation) by the inclusion of a metric degree of freedom on the brane, $N$ which allows manifest reparametrization invariance. Varying this degree of freedom,  $\frac{\delta S_{bubble}}{\delta N}=0$, we find \footnote{We have chosen $\sigma_{dS_{T,F}}=+1$ which is the physically relevant case when $0<R<L_{F}<L_{T}$, with same time orienation inside and outside the bubble, and for small enough positive $\mu$. For an extensive analysis of different cases see \cite{Chernov:2007cm} and references therein.}
\beq
\label{Bubble_effective_action_lapse_variation}
\sqrt{f_{dS_T}+ N^{-2}\dot R^2}-  \sqrt{f_{dS_F} + N^{-2}\dot R^2}- 4\pi \mu G_{N} R=0\,.
\eeq
The above equation is the only independent differential equation, since it turns out that the Euler-Lagrange equation obtained by varying $r$ is simply the $\lambda$ derivative of \eqref{Bubble_effective_action_lapse_variation}. Keeping the lapse in the action is important to obtain the first-order equation \eqref{Bubble_effective_action_lapse_variation} from first principles. Having derived it as an Euler-Lagrange equation, we may now fix the gauge on the brane by setting $N=1$, which identifies $\lambda$ with the proper time $s$ along worldlines of fixed angular coordinates on the brane. Then Eq.~(\ref{Bubble_effective_action_lapse_variation}) matches the junction condition one would obtain by variation of the fields at ${\cal{T}}$ in~(\ref{EH_action_bubble}).

We now set $N=1$ and continue $s\to e^{i \gamma'}s$. (We use $\gamma'$ instead of $\gamma$ because the proper time is not exactly the same as the coordinate time which was continued in earlier parts of this paper; we will derive the relationship below.) 
The energy is 
\beq
\label{bubble_Hamiltonian_lapse_cont}
E=e^{i\gamma'}\frac{\pi}{G_{N}}\left[ R\sqrt{f_{dS_{F}}+ e^{-2 i\gamma'}\dot R^{2}}- R\sqrt{f_{dS_{T}}+ e^{-2i\gamma'}\dot R^{2}}+ 4\pi G_{N}\mu  R^{2}\right]\,.
\eeq
Setting $N=1$ and $s\to e^{i \gamma'}s$ into Eq.~\eqref{Bubble_effective_action_lapse_variation} we note that the energy vanishes on-shell, indicative of time reparametrization invariance. To find the classical solutions we first note that $R(s)=0$ is a trivial solution with vanishing energy. To find the other solution branch we rearrange \eqref{Bubble_effective_action_lapse_variation} into the form
\beq
\label{energy_equation_bubble}
e^{-2i\gamma'}\left(\frac{dR}{ds}\right)^{2}+ V(R)=0\,,
\eeq
where $V(R)= 1- \alpha^2 R^2$, and $
\alpha^{2} = L_{F}^{-2}+ \frac{\left(L_{F}^{-2}- L_{T}^{-2}- 16 \pi^{2} G_{N}^2 \mu^2 \right)^2}{64\pi^{2} G_{N}^{2}\mu^2}$. The solution with vanishing energy at $s_{0}$ is
\beq
\label{Bubble_position_proper_time}
R(s)=\alpha^{-1}\text{cosh}\left[\alpha\, e^{i\gamma'}(s -s_{0})+ i(\pi/2)\right]\,.
\eeq
The initial proper time $s_{0}= \frac{\pi}{2}\alpha^{-1}\csc(\gamma')$ is the largest negative time for which $\text{Im}\,R(0)=0$. The final position at $s=0$ is $R(0)= \alpha^{-1}\text{cosh}((\pi/2)\text{cot}\gamma')$ which falls withing the classical allowed region $R>\alpha^{-1}$ for any $\gamma'$. Similar to the models of Sec.~\ref{secgamma} the real time energy vanishes for the final position and momenta, and final states with different $\gamma'$ are connected to each other at the classical level, by real time evolution. The on-shell action is given by the following integral $-i \int_{T_{i}}^{T_{f}}d\tau\,L_{c}=-i e^{i\gamma} \int _{0}^{R(0)} P(R) dR$, where $P(R)$ is the on-shell momenta
\beq
P(R)= \frac{1}{G_{N}}R \ln \left[\frac{\sqrt{1- L_{T}^{-2}R^{2}}}{\sqrt{1- L_{F}^{-2}R^{2}}}~ \left(\frac{\sqrt{\alpha^{2}R^{2}-1}+ R \sqrt{\alpha^{2}-L_{F}^{-2}}}{\sqrt{\alpha^{2}R^{2}-1}+ R \sqrt{\alpha^{2}-L_{T}^{-2}}}\right)\right]\,.
\eeq

As one might expect the bubble action in Eq.~\eqref{Bubble_effective_action_lapse} reduces to Eq.~\eqref{model} with $p=3$ to leading order in $R^{2}/G_{N}\to 0$ limit \cite{1996PhyEs...9..556A}. If we set $N=1$ and expand \eqref{Bubble_effective_action_lapse} to leading order in $G_{N}$ we find the following action
\beq
\label{leading_GN_Bubble_effective_action}
S^{\text{eff}}_{\text{bubble}}= \int ds \left(\frac{4}{3}\pi R^{3}(\epsilon_{F}-\epsilon_{T})\sqrt{1+ \dot R^{2}}- 4\pi                                          \mu R^{2}\right)\,,
\eeq
where $\Lambda_{T,F}=3L_{T,F}^{-2}= 8\pi G_{N}\epsilon_{T,F}$ and $\dot R$ means derivation with respect to proper time $s$. To match with \eqref{Bubble_effective_action_lapse} we also need to express the action in terms of the Lorentzian time $t$. To leading order in $G_{N}$ the relation between $s$and $t$ is usual time dilation relation $dt\approx ds\sqrt{1+ \left(\frac{dR}{ds}\right)^{2}}$. In terms of Lorentzian time $t$ the  Lagrangian of \eqref{leading_GN_Bubble_effective_action} is proportional to \eqref{Bubble_effective_action_lapse} with $p=3$ and $R_{0}= \frac{3\mu}{\epsilon_{F}-\epsilon_{T}}\approx \alpha^{-1}$. 

Similarly we can recover the solution \eqref{Rcirclegeneral} from \eqref{Bubble_position_proper_time} in the $R^2/G_{N} \to 0$ limit. If we continue the time dilation relation we find
\beq
\label{time_dilation_cont}
e^{i\gamma}d\tau\approx e^{i\gamma'}ds\sqrt{1+ e^{-2i\gamma'}\left(\frac{dR}{ds}\right)^{2}}\,.
\eeq
 If we now integrate the time dilation relation, use $\alpha^{-1}\approx R_{0}$ to leading order and the initial conditions $s_{0}$,$\tau_{0}$ we find
\beq
\label{coordinate_proper_time_leading_order}
e^{i\gamma}\tau- R_{0}\text{cot}\gamma= \alpha^{-1}\text{sinh}\left(\alpha e^{i \gamma' }s - \frac{\pi}{2}\text{cot }\gamma'\right)\,.
\eeq
Making use of \eqref{coordinate_proper_time_leading_order} it is easy to check that the solution \eqref{Bubble_position_proper_time} reduces to \eqref{Rcirclegeneral}. The relation between the coordinate time continuation parameter $\gamma$ and the proper time  parameter $\gamma'$ can also be found, if we set $\tau=s=0$ into \eqref{coordinate_proper_time_leading_order}. We find $\text{cot}\gamma=\text{sinh}\left(\frac{\pi}{2}\text{cot}\gamma'\right)$, which together with \eqref{coordinate_proper_time_leading_order} implies $R(0)=R_{0}|\csc\gamma|$ as expected.

Now let us include some time dependence. 
A field theory model that can be mapped more or less straightforwardly to a time-dependent generalization of (\ref{Bubble_effective_action_lapse}) is given by 
\begin{align}
    V= \frac{1}{2}[\lambda_0+(g/a) \chi](\phi^2-a^2)^2 - (\epsilon/2a) \phi + V_0+V(\chi). 
    \label{eq:phichigravity}
\end{align} 
As in the nongravitational example, and for similar reasons, we consider the hierarchy $\lambda_0 a^4 \gg V_0 \gg (V(\chi), \epsilon) >0$, so that the bubble profile is dominated by the $\phi^4$ potential, and the state of the gravitational fields in the true and false vacua can be approximated by de Sitter space. We assume $\chi$ and $\phi$ are weakly coupled, which means both that $g$ is small and that the metric in the bubble background does not substantially alter the background solution for $\chi$, which is consistent with the small $\epsilon$ approximation.

The relevant rolling solution for $\chi$ at $\calO(g^0)$ is homogeneous in the flat slicing
of dS. The static slicing is more convenient for the present problem, so $\chi$ is a function of the form $\chi(t+L \log\sqrt{1-r^2/L^2 })$, with $L^2\approx 3M_p^2/V_0$ and $r$ and $t$ the static radial and time coordinates, respectively. 

This model is obviously not the most general case, since the background spacetimes are approximated as time-independent and the underlying field theory model is certainly fine-tuned. Its main virtue is that the effective action is almost entirely recycled from the time-independent case.
In the thin-wall limit the bubble dynamics of the model~(\ref{eq:phichigravity}) maps to a membrane effective action of the form (\ref{Bubble_effective_action_lapse}) with $\mu\rightarrow \mu(t_{dS_F},R)$. The spacetime dependence of the tension appears at $\calO(g)$ and is inherited from the background solution for $\chi$ in static coordinates.

If we set lapse variation of \eqref{Bubble_effective_action_lapse} to zero with $\mu$ time dependent, rearrange and apply a $\lambda$ derivative on both sides we find
\begin{equation} \label{grmembraneeq2}
 \frac{d}{d\lambda}\left[\frac{R}{\left(\frac{\pa q}{\pa N}\right)}\left(2\sqrt{f_{dS_{T}}+ N^{-2}\dot R^2}- 2\sqrt{f_{dS_{F}}+ N^{-2}\dot R^{2}} -\tilde{\mu} R\right)\right]= -N R^{2}\frac{\pa \tilde{\mu}}{\pa t_{dS_{F}}},
 \end{equation}
 where we defined $q:= \sqrt{f^{-1}_{dS_{F}}N^{2}+ f^{-2}_{dS_{F}}\dot R^{2}}$ and $\tilde{\mu}:= 8\pi G_{N}\mu$.
Eq.~(\ref{Bubble_effective_action_lapse_variation}) is a first order equation analogous to a conservation of energy law. Eq.~(\ref{grmembraneeq2}) is a second order equation describing the non-conservation of energy by the varying $\mu$.

From this point one may proceed to rotate the time contour and seek suitable tunneling solutions connecting the vacuum to nucleated bubble states. Similarly to the nongravitational cases, at leading order (in the derivative of $\mu$ with respect to the flat time slicing) corrections to the tunneling probability may be found from 
the unperturbed solutions \eqref{Bubble_position_proper_time} and extremization over $\gamma$.

\section{An instanton in a Kaluza-Klein cosmology}
\label{sec:KK}
We conclude with an example where an exact solution of the field equations in a time-dependent background may be obtained. The setting is purely gravitational and describes tunneling in a certain Kaluza-Klein cosmology. We begin with the metric
\begin{align}
ds^2=-dy^2 + y^2 d\phi^2 + dt^2+dx^2 + x^2 d\psi^2,
\label{classicalMilne}
\end{align}
with timelike coordinate $y$ and  non-standard Kaluza-Klein identification
\begin{align}
(t,\phi)\sim (t+2\pi n \frac{\mu}{\sqrt{\mu-a^2}},\phi - 2\pi n  \frac{a}{\sqrt{\mu-a^2}})
\label{MPperiodicity}
\end{align}
and $0\leq a^2\leq\mu$. For nonzero $a$ the KK circle is twisted with spatial Milne-type coordinate $\phi$. Indeed, the metric~(\ref{classicalMilne}) can be extended in a similar way to the Milne extension to Minkowski. First we define
\begin{align}
    \tilde\phi = \phi + (a/\mu) t.
\end{align} 
At fixed $\tilde\phi$ the periodicity is simply $t\sim t+2\pi n \frac{\mu}{\sqrt{\mu-a^2}}$. We then transform $y^2=z^2-w^2$, $\tanh \tilde\phi =z/w$ to obtain
\begin{align}
ds^2=-dz^2+\left(1+\frac{a^2}{\mu^2}(z^2-w^2)\right)dt^2+\frac{2a}{\mu}dt(wdz-zdw)+dw^2+dx^2+x^2d\psi^2.
\label{fullvacuum}
\end{align}
This will play the role of our vacuum spacetime. Locally, it is just flat space, but globally there is a KK circle with interesting properties.  
For $a=0$, (\ref{fullvacuum}) is ordinary KK spacetime which can decay by nucleating a bubble of nothing~\cite{wittenBON}. The instanton is the Euclidean continuation of 5D Schwarzschild-Tangherlini. For nonzero $a$, the vacuum spacetime is more exotic: it is cylindrically symmetric, time dependent, and exhibits a Killing horizon for $\partial_t$ at $w^2=w_{s}^2\equiv z^2+\mu^2/a^2$, beyond which the KK circles are closed timelike curves (CTCs). The presences of CTCs is certainly an upsetting feature, and we will assume   that (\ref{fullvacuum}) is only valid out to some cutoff $\bar w^2$, with
\begin{align}
\bar w^2 < w_s^2,
\label{eq:barw}
\end{align} 
beyond which the spacetime is  ``completed" in a more physical way. (This is clearly a strong assumption. We will see at least a physical argument below that for $a^2\ll\mu$ there is a parametric separation between the spacetime region supporting bubble nucleation and the region exhibiting CTCs, lending some support to the idea that the regions can be studied independently.)   
The most interesting feature is that for $w^2<w_s^2$ the proper circumference of the KK circle is  time $(z)$ dependent, shrinking to a minimum size $2\pi\sqrt{\frac{\mu^2-a^2 w^2}{\mu-a^2}}$ at $z=0$. 
Then an explicit instanton solution describing the decay of this approximate spacetime may be obtained by  analytic continuation of a Myers-Perry black hole~\cite{MP}.\footnote{Ref.~\cite{dowkeretal} also studied continuations of Myers-Perry solutions. We will differ in the choice of  continuation, and therefore also in the interpretation.}

The 5D Myers-Perry solution with one nonzero angular momentum parameter  is given in Boyer-Lindquist-type coordinates by the Lorentzian metric
\begin{align}
\begin{split}
ds^2 =& -dt^2 +\sin^2\theta(r^2+a^2) d\phi^2 +\frac{\mu}{\rho^2}(dt+a\sin^2\theta d\phi)^2 \\& +\frac{\rho^2}{r^2+a^2-\mu}dr^2 +\rho^2 d\theta^2 + r^2\cos^2\theta d\psi^2\,,
\end{split}
\label{MP1J}
\end{align}
where $\rho^2 = r^2+a^2\cos^2\theta$, $a^2<\mu$, and the horizon is at $r_H=\sqrt{\mu-a^2}$. Let us review the limiting behaviors. Asymptotically, the metric takes the form
\begin{align}
ds^2=-dt^2 + dr^2 +r^2 (d\theta^2 +\sin^2\theta d\phi^2 + \cos^2\theta d\psi^2).
\end{align}
The angular part describes an $S^3$  in ``Hopf" coordinates. $\phi,\psi$ range from 0 to $2\pi$ and $\theta$ ranges from 0 to $\pi/2$. Next we inspect the near-horizon limit, following the notation of~\cite{dowkeretal}. Define the Killing vector $\ell = \partial_t -(a/\mu) \partial_\phi$. Then the horizon is a fixed surface of $\ell$, $\ell^2|_{r_H}=0$. Introducing the coordinate $\tilde\phi=\phi + (a/\mu) t$ as above, 
which is constant on the orbits of $\ell$, the near-horizon metric at fixed $\theta,\tilde\phi,\psi$ takes the form
\begin{align}
ds^2=2\left(\mu-a^2\sin^2\theta\right)\left(-\frac{r_H}{\mu^2}(r-r_H)dt^2+\frac{1}{4r_H}\frac{dr^2}{r-r_H}\right),
\label{trmetric}
\end{align}
and in terms of a proper radial coordinate $v$  it is proportional to
\begin{align}
ds^2=dv^2-v^2\left(\frac{\mu-a^2}{\mu^2}\right)dt^2.
\label{nearhorizonMPvt}
\end{align}
Now we perform the analytic continuation
\begin{align}
t\rightarrow it,~~\phi \rightarrow i\phi,~~\theta\rightarrow i\theta.
\label{MPcontinuation3}
\end{align}
Under this map $\theta$ and $\phi$ become noncompact, and $\theta$ becomes timelike. 
The continued metric is
\begin{align}
ds^2 = \frac{r^2+a^2\cosh^2\theta}{r^2+a^2-\mu}&dr^2+dt^2 -\frac{\mu}{r^2+a^2\cosh^2\theta}(dt-a\sinh^2\theta d\phi)^2
 \nonumber\\
 + r^2&\left(-\left[1+\frac{a^2\cosh^2\theta}{r^2}\right] d\theta^2+\sinh^2\theta\left[1+\frac{a^2}{r^2}\right] d\phi^2 + \cosh^2\theta d\psi^2\right).
 \label{fullMPcontinuation3}
\end{align}
Asymptotically in $r$ it behaves as
\begin{align}
ds^2= dr^2 + dt^2  + r^2(-d\theta^2 + \sinh^2\theta d\phi^2 + \cosh^2\theta d\psi^2).
\end{align}
Transforming  $\tanh^2\theta=(z^2-w^2)/x^2,~r^2=x^2-z^2+w^2,~\tanh \tilde\phi =z/w$, we recover the vacuum metric~(\ref{fullvacuum}). The periodicities also match, as can be seen from the  continuation of the near-horizon metric~(\ref{nearhorizonMPvt}). Absence of a conical singularity at $r=r_H$ implies the periodicity $t\sim t+2\pi\frac{\mu}{\sqrt{\mu-a^2}}$ at fixed $\tilde\phi$. Then in terms of the original coordinates we  recover  precisely the periodicity~(\ref{MPperiodicity}). We also conclude that the radial coordinate is bounded, $r\geq r_H$, capping off smoothly at $r=r_H$.

Thus the spacetime~(\ref{fullMPcontinuation3}) describes something bubble-like embedded in the cosmology (\ref{fullvacuum}). 
The induced metric on the bubble wall worldvolume $ r\rightarrow r_H$ is
\begin{align}
ds^2=-(\mu+a^2\sinh^2\theta)d\theta^2+(\mu-a^2)\cosh^2\theta d\psi^2 + \frac{\mu^2\sinh^2\theta}{\mu+a^2\sinh^2\theta}d\tilde\phi^2
\end{align}
which is somewhat opaque. On the wall we make the coordinate transformation
\begin{align}
\sinh\tau&=\sinh\theta\cosh\tilde\phi\nonumber\\
\cosh\tau\cos\chi &= \sinh\theta\sinh\tilde\phi.
\label{hopf2sphere}
\end{align}
$\tau$ is a timelike coordinate on the wall, and the induced metric at fixed $\tau$ becomes, to $\calO(a^2)$,
\begin{align}
ds^2= \cosh ^2\tau\left[\left(\mu+ \frac{a^2}{4}  \left((\cos 2\chi-1) \cosh 2
   \tau - (3 \cos 2\chi+1)\right) \right)d\chi^2+ \left(\mu
   -a^2\right) \sin ^2(\chi )d\psi^2\right].
\end{align}
At $\tau=0$ it
 has the form of the induced metric on a prolate spheroid embedded in $\mathbb{R}^3$, in spherical coordinates with polar angle $\chi$ and azimuthal angle $\psi$, and ratio of semi-axis lengths $c/b=1/\sqrt{1-a^2/\mu}$. It expands exponentially and deforms as time evolves.

To summarize, after the continuation~(\ref{MPcontinuation3}), the 5D black hole becomes a Lorentzian geometry corresponding to an expanding bubble, embedded in a KK space with nonstandard identification of the KK circle. It is therefore a candidate decay product for the classical solution~(\ref{fullvacuum}). More precisely, since we assume the solution~(\ref{fullvacuum}) is only valid for $w^2<\bar w^2\ll (\mu/a)^2$, we have a candidate decay product so long as the bubble fits, $r_H^2 \ll \bar w^2$. For small $a$ the cutoff must fall between $\mu \ll \bar w^2 < (\mu/a)^2$.

 Is there a saddle point of the gravitational action with action $S$, such that the leading order decay amplitude is $e^{-Re(S)}$? 

Because the black hole is stationary instead of static, there is no real Euclidean section that can be obtained by analytic continuation of the coordinates alone. We  may obtain a suitable candidate from~(\ref{MP1J})  by making the $\tilde\phi=\phi + (a/\mu) t$ substitution and continuing only $t\rightarrow it$. The result is a ``quasi-Euclidean" metric with a bubble wall at $r=r_H$ and the Euclidean vacuum, Eq.~(\ref{fullvacuum}) with $z\rightarrow -iz$, at large radius $\sqrt{x^2+w^2}$. The induced metric on the $z=0$ nucleation hypersurface agrees between the quasi-Euclidean instanton and the Lorentzian nucleated bubble~\ref{fullMPcontinuation3}.

The action of the quasi-Euclidean solution is
\begin{align}
    S=\frac{\pi ^2 \mu ^2}{G_5 \sqrt{\mu -a^2}}\,.
    \end{align}
It recovers the action of Witten's bubble in the limit $a\rightarrow 0$ and diverges in the extremal limit $a\rightarrow \sqrt{\mu}$, where the KK circle size diverges.

The quasi-Euclidean solution describes the amplitude to evolve from a state with no bubble at some early time $z_i$ to a state with a bubble at some late time $z_f$. The action controls the amplitude for the bubble to nucleate, and  the analytic continuation describes the subsequent bubble evolution. Since the background is time dependent there is no time translation symmetry and the instanton computes a probability rather than a rate. Note that the transition occurs when the KK circle is smallest near the origin of the spatial coordinates. This corresponds to the smallest potential barrier.

We conclude this section with a few comments.
\begin{itemize}
    \item The bubble of nothing example studied in this section is similar to previous examples we have seen, in that it involves a saddle point off in the complexified phase space. Quasi-euclidean metrics based on the continuation of Kerr-like solutions are known to play a variety of useful other roles~\cite{Monteiro:2009tc,Witten:2021nzp}.
    \item In a few other ways, the present example is different. First, the presence of a time reversal symmetry means there is a time when the quasi-Euclidean solution returns to the real axis with zero momentum. (Indeed, in the notation of previous sections we have only considered the Euclidean continuation $\gamma=-\pi/2$.) Also, we have only found an explicit solution when the initial time is early, so that the KK circle is shrinking, and the final time is late, when it is expanding again. In previous examples we were able to find families of instantons for more general initial and final times. Finally, the bubble is neither thin wall nor spherically symmetric. In all other cases, the background was rotationally invariant and the bubble was thin-wall and spherical by assumption.
    \item At first sight, a more natural starting point would be the background~(\ref{classicalMilne}) with standard periodicity $\phi\sim\phi + 2\pi$.  However, this spacetime is singular at $y=0$, where the circle shrinks to zero size. The instanton in this case is also singular and has zero action -- there is no barrier to tunneling at this moment. The purpose of the twisted periodicity is to regulate the circle so that it is never smaller than $\mu/\sqrt{\mu-a^2}$ at the origin. 
\end{itemize}

\section{Discussion}
\label{sec:discussion}

We have surveyed the effects of slow explicit time dependence on barrier penetration and bubble nucleation in some simple models.  The problem is interesting not because the  effects are typically dramatic, but because the usual semiclassical techniques require some development in order to calculate the effects. It appears to be useful generalize the Wick rotation to other contours in the complex time plane, which provides semiclassical access to different tunneling times during a fixed time window. ``Nucleation," in the sense of the bubble in the final state, need not occur at rest, and  semiclassical trajectories with nonzero final momentum  can be interpreted as providing a saddle point solution for a path integral with wavepacket initial and final states, with the real and imaginary parts of the final momentum mapping to phases and gradients of the wavepacket.

Reduction to  effective quantum mechanical models simplified the problem enormously, essentially from a nonlinear PDE back to a nonlinear ODE, as one has in the fully field-theoretic treatment of $O(4)$-symmetric problems. It is unclear if there is a practical generalization of our methods to full field theories, although in some favorable cases exact solutions can be guessed.

We conclude with a few directions for
future work.  First, our analysis was entirely at the leading semiclassical order. The fluctuation determinant has interesting technical features, requiring the solution of flow equations to identify suitable middle-dimensional integration cycles~\cite{Witten:2010zr}. Although we have seen a variety of physical features that suggest the complex trajectories studied here indeed compute meaningful tunneling probabilities, explicit computation of the fluctuation determinant in a model would provide further support for the interpretation.

Second, in the effective model of membrane nucleation coupled to gravity, the construction of the effective action itself is somewhat nontrivial, and we only illustrated the construction in some relatively simple classes of models. Further analysis and generalization could be of interest for cosmological applications. 

Finally, our study of exact solutions was  limited to a single example.  Some continuations of other black hole and black ring solutions in higher dimensions presumably describe bubble of nothing decays in more exotic Kaluza-Klein cosmologies, and it would be interesting to catalog them.

\section*{Acknowledgements}
We thank Szilard Farkas, Adam Nahum, LianTao Wang, and Yikun Wang for useful discussions.  
This work was supported in part by the U.S.
Department of Energy, Office of Science, Office of High Energy Physics under award number DE-SC0015655.

\bibliographystyle{utphys}
\bibliography{tunneling_arxiv}
 
\end{document}